\documentclass{article}

\usepackage{arxiv}

\usepackage[utf8]{inputenc} % allow utf-8 input
\usepackage[T1]{fontenc}    % use 8-bit T1 fonts
\usepackage[hidelinks]{hyperref}      % hyperlinks
\usepackage{url}            % simple URL typesetting
\usepackage{booktabs}       % professional-quality tables
\usepackage{amsfonts}       % blackboard math symbols
\usepackage{nicefrac}       % compact symbols for 1/2, etc.
\usepackage{microtype}      % microtypography
\usepackage{lipsum}
\usepackage{graphicx}

\usepackage{amsmath,mathrsfs,enumerate,multirow,appendix}
\usepackage{longtable, caption}
\usepackage{amsfonts, amssymb}
\usepackage{xcolor}
\usepackage{placeins}

\usepackage[table]{xcolor}
\definecolor{Color1}{HTML}{D0D0D0}   % gray
\definecolor{Color2}{HTML}{C7DFEC}   % blue-gray
\definecolor{Color3}{HTML}{CCE6C4}   % green-gray

\usepackage[round]{natbib}

\usepackage{bm}

\makeatletter
\renewcommand\@biblabel[1]{#1.}
\makeatother

\newcommand{\ksup}{\ensuremath{^{(k)}}}
\newcommand{\kosup}{\ensuremath{^{(k_1)}}}
\newcommand{\ktsup}{\ensuremath{^{(k_2)}}}
\newcommand{\ssup}{\ensuremath{^{(s)}}}

\graphicspath{ {./images/} }

\title{Integrating Prioritized and Non-Prioritized Structures in Win Statistics}

\author{
    Yunhan Mou \\
    Department of Biostatistics,\\
    Yale School of Public Health\\
    New Haven, Connecticut, USA \\
    \texttt{yunhan.mou@yale.edu} \\
    \And
    Scott Hummel \\
    Frankel Cardiovascular Center,\\
    University of Michigan\\
    Ann Arbor, Michigan, USA;\\
    Department of Internal Medicine, \\Division of Cardiovascular Medicine,\\
    VA Ann Arbor Health System\\
    Ann Arbor, Michigan, USA\\
    \texttt{scothumm@umich.edu} \\
    \And
    Yuan Huang\thanks{Corresponding authors}\\
    Department of Biostatistics,\\
    Yale Center for Analytical Sciences,\\
    Yale School of Public Health\\
    New Haven, Connecticut, USA; \\
    \texttt{yuan.huang@yale.edu}\\
}

\begin{document}
\maketitle
% \vspace{-0.1in}
\begin{abstract}
Composite endpoints are frequently used as primary or secondary analyses in cardiovascular clinical trials to increase clinical relevance and statistical efficiency. Alternatively, the Win Ratio (WR) and other Win Statistics (WS) analyses rely on a strict hierarchical ordering of endpoints, assigning higher priority to clinically important endpoints. However, determining a definitive endpoint hierarchy can be challenging and may not adequately reflect situations where endpoints have comparable importance. 
In this study, we discuss the challenges of endpoint prioritization, underscore its critical role in WS analyses, and propose Rotation WR (RWR), a hybrid prioritization framework that integrates both prioritized and non-prioritized structures. 
By permitting blocks of equally-prioritized endpoints, RWR accommodates endpoints of equal or near equal clinical importance, recurrent events, and contexts requiring individualized shared decision making. Statistical inference for RWR is developed using U-statistics theory, including the hypothesis testing procedure and confidence interval construction. Extensions to two additional WS measures, Rotation Net Benefit and Rotation Win Odds, are also provided.
Through extensive simulation studies involving multiple time-to-event endpoints, including recurrent events, we demonstrate that RWR achieves valid type I error control, desirable statistical power, and accurate confidence interval coverage. We illustrate both the methodological and practical insights of our work in a case study on endpoint prioritization with the SPRINT clinical trial, highlighting its implications for real-world clinical trial studies.
\end{abstract}

% keywords can be removed

\section{Introduction} \label{intro}

Composite endpoints are widely adopted in cardiovascular clinical trials to enhance clinical relevance and improve statistical efficiency. A common example is the combination of mortality and hospitalization, which allows investigators to capture a broader spectrum of treatment effects than within a single primary endpoint. As an alternative to the traditional time-to-first-event analysis, which fails to address the differential clinical importance of endpoints, the Win Ratio (WR) method \citep{pocock2012win} and more general frameworks of Win Statistics (WS) \citep{dong2023win} or Generalized Pairwise Comparison \citep{buyse2010generalized, verbeeck2023generalized} have gained substantial attention in recent years. These approaches perform pairwise comparisons between patients across multiple endpoints, ordered by a strict hierarchy that prioritizes clinically important events. For instance, death is typically placed at the top of the hierarchy, followed by hospitalization. This prioritization structure aligns the statistical analysis with clinical priorities.
Given its advantages, WR has increasingly been recognized as a mainstream approach for composite endpoints \citep{gasparyan2021adjusted}. Clinical trials, including EMPULSE (NCT0415775), VIP-ACS (NCT04001504), and DAPA-HF (NCT03036124), have prespecified WR as their primary outcome measure.

As the use of WR has expanded in clinical applications, methodological research has also grown substantially. Stratified WR was introduced to account for discrete baseline variables \citep{dong2018stratified, gasparyan2021adjusted}. Statistical inference procedures for WR, including hypothesis testing and confidence interval, have been extensively investigated \citep{luo2015alternative, bebu2016large, dong2016generalized, mao2019alternative}. \citet{oakes2016win} examined the estimation of the WR under censoring using integration-based expressions. In settings with potentially covariate-dependent right censoring, inverse probability of censoring–weighted approaches have been proposed to remove dependence on the censoring distribution \citep{dong2020inverse, dong2021adjusting}. \citet{mao2024defining} explored the underlying estimand of WR. 
Within the broader family of WS \citep{dong2023win, verbeeck2023generalized, buyse2025handbook}, other effect measures that share the endpoint prioritization principle are available, including the net benefit \citep{buyse2010generalized}, win loss \citep{luo2017weighted}, win probability \citep{gasparyan2021adjusted}, win odds \citep{dong2020winodds}, event-specific WR \citep{yang2021eventct}, and beyond.

A design feature of typical WS is the requirement to specify a complete hierarchy of endpoints. Endpoints must be strictly ordered according to their clinical importance, with no provision for assigning equal ranks. While ensuring priority of endpoints with higher importance, this design also imposes rigidity. To the best of our knowledge, within WS, the only setting in which endpoints can be considered on an equal priority is through non-prioritized pairwise comparison \citep{verbeeck2019generalized} or event-specific WR \citep{yang2021eventct} with equal weights, where all endpoints contribute symmetrically to the net treatment benefit without hierarchical ordering. However, these approaches necessarily remove the hierarchical structure and do not allow prioritization and non-prioritization of endpoints to coexist within the same structure. 
In practice, however, investigators may find it challenging to impose a complete hierarchy across all endpoints (i.e., full prioritization), while treating all endpoints as of equal importance (i.e., non-prioritization) remains unsatisfactory. We hereby highlight three scenarios in which an intermediate approach between full prioritization and non-prioritization may be particularly appealing.

First, certain endpoints differ markedly in clinical importance, whereas others are relatively comparable. For instance, in a composite endpoint consisting of both fatal and non-fatal events, death is typically regarded as the most serious outcome and is prioritized above all others. By contrast, non-fatal components of the 3-point major adverse cardiovascular events (MACE) composite \citep{bosco2021major} such as stroke and myocardial infarction may have comparable clinical weights in some situations, rendering a strict ordering among them unnecessary.
Second, for non-fatal endpoints that may recur, restricting attention to the time of the first event could overlook important clinical information contained in subsequent episodes. For instance, a reanalysis of the DAPA-HF trial (NCT03036124) showed that dapagliflozin substantially reduced the total number of worsening heart failure events, whereas its effect on the conventional time-to-first-event endpoint was less pronounced \citep{jhund2021dapagliflozin}. Moreover, accounting for recurrent events has been shown to improve statistical efficiency \citep{claggett2018comparison}. Within the WS framework, a straightforward way to incorporate recurrent event data is by including the cumulative frequency as a layer \citep{mao2022recurrentevent, maurer2018tafamidis}. Additional summaries, such as the time to first or time to last episode, may still be considered simultaneously to capture different aspects of disease progression. However, it is not always necessary to impose a strict prioritization among these summaries, as they reflect complementary facets of the same underlying endpoint.
Third, shared decision making (SDM) represents an additional scenario.
The WS methods are potentially well suited for SDM because they are conceptually intuitive and can incorporate patient-specific preferences in the prioritization of endpoints \citep{salvaggio2025generalized}. Unlike conventional trial analyses that prioritize endpoints based on scientific priors, SDM emphasizes that patients may value endpoints differently depending on their personal circumstances and goals. For example, while physicians may uniformly regard death as the most important outcome, some patients may place greater emphasis on functional status, quality of life, or avoidance of hospitalization \citep{stanek2000preferences, stevenson2008changing, spertus2019integrating}. In practice, this patient centered perspective implies that some individuals may choose to assign equal weight to certain endpoints, even when they differ in conventional clinical importance. Allowing for such equal priority of endpoints acknowledges the diversity of patient preferences and may improve the alignment of trial evidence with patient values.

The above scenarios motivate us to propose the Rotation Win Ratio (RWR), featuring a hybrid prioritization strategy that extends WR within the WS framework. The key idea of RWR is to preserve prioritization while permitting equal priority among endpoints as needed. Conceptually, this is achieved by constructing a set of “rotations” of the endpoint ordering whenever a group of endpoints is intended to be analyzed with equal priority. For example, suppose death is clearly prioritized above all other endpoints, while myocardial infarction and stroke are considered of similar importance but more important than hospitalization. Rather than forcing a single strict sequence, RWR rotates the order of myocardial infarction and stroke, performing pairwise comparisons under both sequences and then integrating them as a single RWR value, an effect measure similar to the standard WR. In doing so, endpoints that are clearly prioritized retain their hierarchical position, whereas endpoints deemed equal attention are treated symmetrically. By blending the full prioritization and non-prioritization into a single structure, RWR enables a more flexible representation of clinical and patient centered endpoint ordering, while retaining the interpretability and rigor of the standard WR. Using U-statistics theory, we develop the hypothesis testing procedure and closed-form confidence interval construction for RWR.
Our illustration will focus primarily on the WR measure (i.e., RWR), while also introducing the methodology for the net benefit and win odds measures, two other widely used effect measures within the WS framework.
To evaluate the operating characteristics of RWR, we conduct extensive simulation studies under settings designed to mimic realistic composite endpoints in cardiovascular trials. In the first setting, a fatal event is placed at the top of the hierarchy, followed by three non-fatal time-to-event endpoints of comparable clinical importance. In the second setting, a fatal event is prioritized above a recurrent non-fatal endpoint, where the recurrent non-fatal endpoint is summarized using the number of recurrences, the time to the first recurrence, and the time to the last recurrence.
For both settings, we vary key design and data generating features to present a comprehensive evaluation of operating characteristics, including type I error, power, and confidence interval coverage.
We further illustrate the proposed method using data from the Systolic Blood Pressure Intervention Trial (SPRINT), which features multiple clinically relevant endpoints of varying importance. In this case study, we also demonstrate how to summarize and report layer-level proportions of wins, losses, and ties under the hybrid-prioritization structure, thereby aligning the RWR analysis with the common reporting standards for the WS.

\section{Statistical Methods}\label{method}

In this section, we first review the standard WR method and then present the proposed RWR. To fix the notation, we assume a clinical trial with $N$ participants and two treatment arms with $N_t$ participants being in the treatment group and $N_c$ participants being in the control group. 
For participant $i$, let $\bm Y_i$  denote the observed outcome vector across the $q$ endpoints, with superscripts $t$ and $c$ indicating the treatment and control groups, respectively.

\subsection{Standard Win Ratio: A review}\label{WR:review}

The standard WR is based on pairwise comparisons across study participants. For the comparison between participants $i$ and $j$ from contrasting treatment groups, we define the winning function $W(\bm Y_i^t, \bm Y_j^c) = 1$ if $i$ has a more favorable outcome (win) than $j$, and $W(\bm Y_i^t, \bm Y_j^c) = 0$ if $i$ has a less favorable outcome (loss) than $j$ or if neither is more favorable hence the comparison is uninformative or indeterminate (tie). Analogously, we define the losing function $L(\bm Y_i^t, \bm Y_j^c) = 1$ if $i$ has a less favorable outcome than $j$, and $L(\bm Y_i^t, \bm Y_j^c) = 0$ otherwise. Within each pairwise comparison, to determine the comparison result, one examines the endpoints following a specified sequence of endpoint comparisons. Lower layer endpoints are considered only when the higher layer comparisons are indeterminate. The WR can be calculated by 
$$
\text{WR} = \frac{\sum_{i=1}^{N_t} \sum_{j=1}^{N_c} W(\bm Y_i^t, \bm Y_j^c)}
                {\sum_{i=1}^{N_t} \sum_{j=1}^{N_c} L(\bm Y_i^t, \bm Y_j^c)}.
$$

\subsection{Rotation Win Ratio}

To extend WR to settings where blocks of endpoints are meant to hold equal priority, we introduce the Rotation Win Ratio (RWR). 
Here, each block corresponds to a layer in the fully prioritized analysis. However, because RWR allows multiple equally prioritized endpoints within a single block, rather than having a single endpoint for each layer as in a typical fully prioritized structure, we use the term block instead of layer to avoid ambiguity.
Denote the full set of endpoints by $\mathcal{Y}=\{1,2,\ldots,q\}$ and partition them into ordered blocks, with strict prioritization across blocks and equal priority within each block.
To be specific, we may partition $\mathcal{Y}$ into $R$ disjoint blocks,
\[
\mathcal{Y} =\bigcup_{r=1}^{R} \mathcal{G}_r, 
\quad \mathcal{G}_r \cap \mathcal{G}_{r'} = \varnothing \;\; (r\neq r'), 
\quad \mathcal{G}_1 \succ \mathcal{G}_2 \succ \cdots \succ \mathcal{G}_R ,
\]
where “$\succ$” denotes strict clinical priority \emph{across} blocks, while endpoints \emph{within} each block $\mathcal{G}_r$ are to be treated equally. 
Under the hybrid-prioritization framework in RWR, equal priority is operationalized by rotating the order of endpoints within each block while preserving the order of the blocks. 
Let $\mathfrak{S}(\mathcal{G}_r)$ denote the set of permutations of $\mathcal{G}_r$. 
Define the rotation set as the Cartesian product, we have
$
\mathcal{R}_{\mathrm{rot}} = \mathfrak{S}(\mathcal{G}_1)\times \cdots \times \mathfrak{S}(\mathcal{G}_R),
p = \bigl|\mathcal{R}_{\mathrm{rot}}\bigr| = \prod_{r=1}^{R} \bigl|\mathcal{G}_r\bigr|!,
$
where $|\cdot|$ is the cardinality and $p$ is number of rotations involved.
As an example, for 6 endpoints and partition
$\mathcal{G}_1=\{1\},\mathcal{G}_2=\{2,3\},
\mathcal{G}_3=\{4,5\},\mathcal{G}_4=\{6\}.
$
The $p = 1!\cdot 2!\cdot 2!\cdot 1! = 4$ rotations with possible endpoint orders: 
$
(1 \Vert 2,3 \Vert 4,5 \Vert 6), 
(1 \Vert 2,3 \Vert 5,4 \Vert 6), 
(1 \Vert 3,2 \Vert 4,5 \Vert 6), 
(1 \Vert 3,2 \Vert 5,4 \Vert 6),
$
where $\Vert\ $ is used to partition blocks with distinct priority.

For the $k$-th rotation, we define winning and losing functions $W\ksup$ and $L\ksup$ based on the corresponding order of endpoints. Let $n_t\ksup$ and $n_c\ksup$ be the number of wins and losses in the treatment group at this rotation, which can be counted as:
\begin{equation} \label{cal_nt_nc}
    n_t\ksup = \sum_{i=1}^{N_t} \sum_{j=1}^{N_c} W\ksup\left(\bm Y^t_i, \bm Y^c_j\right), 
    \quad
    n_c\ksup = \sum_{i=1}^{N_t} \sum_{j=1}^{N_c} L\ksup\left(\bm Y^t_i, \bm Y^c_j\right).
\end{equation}
The RWR measure is calculated by
\begin{equation} \label{cal_RWR}
    \text{RWR} = \frac{\sum_{k=1}^p n_t^{(k)}}{\sum_{k=1}^p n_c^{(k)}}.
\end{equation}

The resulting RWR measure is interpreted in the same way as the standard WR, while extending the framework to accommodate a hybrid prioritization structure that combines strict hierarchy across blocks and equal priority of endpoints within blocks.

\subsection{Statistical Inference for Rotation Win Ratio}

In this subsection, we develop inferential procedures for RWR. For the $k$-th rotation ($k=1,2,\ldots,p$), $n_t\ksup$ follows the asymptotic normal (AN) distribution as shown below \citep{wei1985combining, dong2016generalized}:
\begin{eqnarray*}
    n_t\ksup &\sim& AN(N_t N_c \theta_t\ksup,\sigma_{t\ksup}^2), \\   
    \theta_t\ksup &=& \mathbb{E}[W\ksup(\bm Y^t_i,\bm Y^c_j)], \\
    \sigma_{t\ksup}^{2} &=& \frac{1}{N_t}\sigma_{t\ksup1}^{2} + 
                        \frac{1}{N_c}\sigma_{t\ksup2}^{2}, \\
    \sigma_{t\ksup1}^{2} &=& \frac{N_t N_c}{N_c-1}
                    \sum_{i=1}^{N_t}\sum_{j=1}^{N_c}
                    \sum_{\substack{j'=1 \\ j'\neq j}}^{N_c}
                    \left[W\ksup(\bm Y^t_i,\bm Y^c_j)-\theta_t\ksup\right]
                    \left[W\ksup(\bm Y^t_i,\bm Y^c_{j'})-\theta_t\ksup\right],\\
    \sigma_{t\ksup2}^{2} &=& \frac{N_t N_c}{N_t-1}
                \sum_{j=1}^{N_c}\sum_{i=1}^{N_t}
                \sum_{\substack{i'=1 \\ i'\neq i}}^{N_t}
                \left[W\ksup(\bm Y^t_i,\bm Y^c_j)-\theta_t\ksup\right]
                \left[W\ksup(\bm Y^t_{i'},\bm Y^c_j)-\theta_t\ksup\right].
\end{eqnarray*}

Analogously, $n_c\ksup \sim AN(N_t N_c \theta_c\ksup,\sigma_{c\ksup}^2)$. Further denote winning and losing functions as $F_{t}\ksup(\cdot,\cdot) = W\ksup(\cdot,\cdot)$ and $F_{c}\ksup(\cdot,\cdot)=L\ksup(\cdot,\cdot)$, respectively, then the covariance of two counts $n_{d_1}\kosup$ and $n_{d_2}\ktsup$, where $d_1, d_2 \in \{t,c\}$ and $k_1,k_2 \in \{1,2,\ldots, p\}$ , can be obtained as: 
\begin{eqnarray*}
    \sigma_{d_1\kosup d_2\ktsup} &=& \frac{1}{N_t}\sigma_{d_1\kosup d_2\ktsup 1} + 
        \frac{1}{N_c}\sigma_{d_1\kosup d_2\ktsup 2}, \\
    \sigma_{d_1\kosup d_2\ktsup 1} &=& \frac{N_t N_c}{N_c-1}
        \sum_{i=1}^{N_t} \sum_{j=1}^{N_c} \sum_{\substack{j'=1 \\ j'\neq j}}^{N_c}
        \left[ F_{d_1}\kosup(\bm Y_i^t, \bm Y_j^c) - \theta_{d_1}\kosup \right]
        \left[ F_{d_2}\ktsup(\bm Y_i^t, \bm Y_{j'}^c) - \theta_{d_2}\ktsup \right], \\
    \sigma_{d_1\kosup d_2\ktsup 2} &=& \frac{N_t N_c}{N_t-1}
        \sum_{j=1}^{N_c} \sum_{i=1}^{N_t} \sum_{\substack{i'=1 \\ i'\neq i}}^{N_t}
        \left[ F_{d_1}\kosup(\bm Y_i^t, \bm Y_j^c) - \theta_{d_1}\kosup \right]
        \left[ F_{d_2}\ktsup(\bm Y_{i'}^t, \bm Y_{j}^c) - \theta_{d_2}\ktsup \right].
\end{eqnarray*}
%With the pairwise comparison results available from the observed dataset, these values can be estimated by replacing the expectations $\theta$ with their estimated values,
Plug-in estimators can be obtained,
e.g., $\hat\theta_{d_1}\kosup = \frac{1}{N_t N_c} \sum_{i=1}^{N_t} \sum_{j=1}^{N_c} \hat F_{d_1}\kosup (\bm Y_i^t, \bm Y_j^c)$, where $\hat F_{d_1}\kosup (\bm Y_i^t, \bm Y_j^c)$ is the observed comparison result.

Therefore, for $\bm n = (n_t^{(1)}, n_t^{(2)}, \ldots, n_t^{(p)}, n_c^{(1)}, n_c^{(2)}, \ldots n_c^{(p)})^\top$, we have:
\begin{eqnarray*}
    \bm n &\sim& AN(N_t N_c \hat{\bm\theta}, \hat{\bm\Sigma}),\\
    \hat{\bm\theta} &=& (\hat\theta_t^{(1)}, \hat\theta_t^{(2)}, \ldots, \hat\theta_t^{(p)}, \hat\theta_c^{(1)}, \hat\theta_c^{(2)}, \ldots, \hat\theta_c^{(p)})^\top,\\
    \hat{\bm\Sigma} &=&  \left[ 
        \widehat{\text{Cov}}(\bm n_i, \bm n_j)
        \right]_{i,j = 1,2,\ldots,p}, 
\end{eqnarray*}
where $\bm n_i$ is the $i$-th coordinate of $\bm n$ and $\widehat{\text{Cov}}(\bm n_i, \bm n_j)$ is the covariance (variance) estimated as shown above. Define transformations $g_t=\left(\bm 1_p^\top,\bm 0_p^\top\right)^\top$ and $g_c=\left(\bm 0_p^\top, \bm 1_p^\top\right)^\top$, where $\bm 0_p, \bm 1_p$ stand for $p-$dimensional vector with all 0 or 1 inputs, respectively. Let $G=\left(g_t^\top, g_c^\top\right)^\top$, then $\tilde{\bm n} = \left(\sum_{k=1}^p n_{t}\ksup, \sum_{k=1}^p n_{c}\ksup\right)^\top = G \bm n$.
By the delta method, we have
\begin{eqnarray} 
    \tilde{\bm n} &\sim& AN\left( N_t N_c G \hat{\bm\theta}, 
                            G \hat{\bm\Sigma} G^\top \right), \label{CI_n_tilde:start}\\
    \log( \tilde{\bm n}) &\sim& AN\left(\log(N_t N_c G \hat{\bm\theta}),
        \hat{\bm\Omega}=\hat D G \hat{\bm\Sigma} G^\top \hat D \right), \\
    \hat D &=& \text{diag}\left\{\frac{1}{N_t N_c g_t^\top \hat{\bm\theta}}, 
        \frac{1}{N_t N_c g_c^\top \hat{\bm\theta}}\right\}. \label{CI_n_tilde:end}
\end{eqnarray}

Therefore, the logarithm of RWR is asymptotically normally distributed as:
\begin{eqnarray*}
    \log(\text{RWR}) &\sim& AN(\log(\hat\mu), \hat\sigma_{\log(\text{RWR})}^2), \\
    \hat\mu &=& \frac{\sum_{k=1}^p \hat\theta_t^{(k)}}{\sum_{k=1}^p \hat\theta_c^{(k)}}, \\
    \hat\sigma_{\log(\text{RWR})}^2 &=& \hat{\bm\Omega}_{11} + 
        \hat{\bm\Omega}_{22} - 2 \hat{\bm\Omega}_{12},
\end{eqnarray*}
where $\hat{\bm\Omega}_{ij}$ is the $(i,j)$-th element of covariance matrix $\hat{\bm\Omega}$. The $(1-\alpha)100\%$ Wald confidence interval for RWR can be calculated accordingly.
To test $H_0: \text{RWR}=1$ (equivalently $ \log(\text{RWR})=0$), null distribution can be obtained by estimating under the null 
\begin{equation*}
    \hat\theta_{t0}\ksup = \hat\theta_{c0}\ksup = \frac{1}{2 N_t N_c} \sum_{i=1}^{N_t} \sum_{j=1}^{N_c} \left[
    \hat W\ksup (\bm Y_i^t, \bm Y_j^c) + \hat L\ksup (\bm Y_i^t, \bm Y_j^c) \right],
\end{equation*}
and substituting in $\hat\theta_{t0}\ksup$ and $\hat\theta_{c0}\ksup$ for the null variance estimator. This completes the specification of the null distribution, and the calculation of standard p-value follows.

\subsection{Stratified Rotation Win Ratio}\label{sec:strat_RWR}

To accommodate clinical trials with the popular stratification design, we extend RWR to its stratified version. Consider a stratified clinical trial with $m$ strata, indexed by $s=1,2,\ldots,m$, with sample sizes $N_t\ssup, N_c\ssup$, and weights $w\ssup$.
Given the independence across strata, pairwise comparisons are conducted within each stratum only. We denote $n_{t}^{(k)(s)}$ and $n_{c}^{(k)(s)}$ as the number of wins and losses in the treatment group at the $k$-th rotation in stratum $s$, accordingly. Similar to the stratification strategy for WR in \citet{dong2018stratified}, the stratified RWR is calculated as:
\begin{equation*}
    \text{RWR}_{\text{stra}} = \frac{\sum_{k=1}^p \sum_{s=1}^m w^{(s)} n_{t}^{(k)(s)}}
    {\sum_{k=1}^p \sum_{s=1}^m w^{(s)} n_{c}^{(k)(s)}}.
\end{equation*}

In the stratified version, analogous to equation (\ref{CI_n_tilde:start}) - (\ref{CI_n_tilde:end}), due to the independence across strata, we have:
\begin{eqnarray*}
    \sum_{s=1}^m w\ssup \tilde{\bm n}\ssup &\sim& 
        AN\left(
        \sum_{s=1}^m w\ssup N_t\ssup N_c\ssup G \hat{\bm\theta}\ssup, 
        \sum_{s=1}^m (w\ssup)^2 G \hat{\bm\Sigma}\ssup G^\top
        \right), \\
    \log \left( \sum_{s=1}^m w\ssup \tilde{\bm n}\ssup \right) &\sim& 
        AN\left(
        \log \left( 
            \sum_{s=1}^m w\ssup N_t\ssup N_c\ssup G \hat{\bm\theta}\ssup \right),
        \hat{\bm\Omega}^{\text{stra}} = 
        \hat D_{\text{stra}} \left[ \sum_{s=1}^m (w\ssup)^2
            G \hat{\bm\Sigma}\ssup G^\top \right]
        \hat D_{\text{stra}}
        \right), \\
    \hat D_{\text{stra}} &=& \text{diag}\left\{ 
        \frac{1}{\sum_{s=1}^m w\ssup N_t\ssup N_c\ssup g_t^\top \hat{\bm\theta}\ssup},
        \frac{1}{\sum_{s=1}^m w\ssup N_t\ssup N_c\ssup g_c^\top \hat{\bm\theta}\ssup}
        \right\}, \\
    \log(\text{RWR}_{\text{stra}}) &\sim& AN(\log(\hat\mu_{\text{stra}}),
        \hat\sigma_{\log(\text{RWR}_{\text{stra}})}^2), \\
    \hat\mu_{\text{stra}} &=& 
        \frac{\sum_{s=1}^m \sum_{k=1}^{p} 
            w\ssup N_t\ssup N_c\ssup \hat\theta_t^{(k)(s)}}
        {\sum_{s=1}^m \sum_{k=1}^{p} 
            w\ssup N_t\ssup N_c\ssup \hat\theta_c^{(k)(s)}}, \\
    \hat\sigma_{\log(\text{RWR}_{\text{stra}})}^2 &=&
        \hat{\bm\Omega}^{\text{stra}}_{11} +
        \hat{\bm\Omega}^{\text{stra}}_{22} - 
        2 \hat{\bm\Omega}^{\text{stra}}_{12}.
\end{eqnarray*}
This gives the Wald confidence interval of $\text{RWR}_{\text{stra}}$ and the stratified RWR test, similar to the non-stratified version.

\subsection{Rotation Win Statistics}
In this subsection, we present the hybrid prioritization framework on two other commonly used WS measures, Net Benefit (NB) \citep{buyse2010generalized} and Win Odds (WO) \citep{dong2020winodds}. Specifically, we propose the methodology of Rotation Net Benefit (RNB) and Rotation Win Odds (RWO), which incorporate the same rotational construction to enable hybrid prioritization across endpoints. Similar to RWR, statistical inference for RNB and RWO, including hypothesis testing and confidence interval construction, is derived using U-statistics theory. Given the shared structure of the derivations, we briefly present the main results below for completeness. Parallel to equation~(\ref{cal_RWR}), let $N_+=\sum_{k=1}^p n_t^{(k)}, N_-=\sum_{k=1}^p n_c^{(k)}$, and $N_0 = pN_tN_c-(N_+ + N_-)$ be the total number of wins, losses, and ties through $p$ rotations, RNB and RWO measures are calculated by:
\begin{equation*}
    \text{RNB} = \frac{N_+ - N_-}{pN_tN_c}, \quad
    \text{RWO} = \frac{N_+ + 0.5N_0}{N_- + 0.5N_0}.
\end{equation*}
For the simplicity of notation, we rewrite equation~(\ref{CI_n_tilde:start}) as 
\begin{equation*}
        \tilde{\bm n} = (N_+, N_-)^\top \sim AN
            \left( \hat {\bm \nu} =N_t N_c G \hat{\bm\theta}, 
                    \hat {\bm \Lambda} = G \hat{\bm\Sigma} G^\top \right).
\end{equation*}
By the delta method, $\text{RNB}$ and $\log(\text{RWO})$ are asymptotically normally distributed with variance \citep{dong2023win}:
\begin{eqnarray*}
    \hat\sigma_{\text{RNB}}^2  &=& 
        \frac{\hat {\bm \Lambda}_{11} + \hat {\bm \Lambda}_{22} - 2\hat {\bm \Lambda}_{12}}
             {\left(pN_tN_c\right)^2},\\
    \hat\sigma_{\log(\text{RWO})}^2 &=& 
        \left(\hat {\bm \Lambda}_{11} + \hat {\bm \Lambda}_{22} - 
            2\hat {\bm \Lambda}_{12}\right) 
        \left( \frac{1}{\eta} + \frac{1}{pN_tN_c - \eta} \right)^2 \Big/4,\\
    \eta &=& \hat {\bm \nu}_1 + 0.5(p N_t N_c - \hat {\bm \nu}_1 - \hat {\bm \nu}_2),
\end{eqnarray*}
where $\hat {\bm \nu}_i$ is the $i$-th coordinate of mean vector $\hat {\bm \nu}$, and $\hat {\bm \Lambda}_{ij}$ is the $(i,j)$-th element of covariance matrix $\hat {\bm \Lambda}$.
This gives the confidence intervals of $\text{RNB}$ and $\text{RWO}$. The hypothesis testing follows that for $\text{RWR}$ by employing $\hat\theta_{t0}\ksup$ and $\hat\theta_{c0}\ksup$ in calculation. The stratified version of $\text{RNB}$ and $\text{RWO}$ can also be similarly conducted as in Section~\ref{sec:strat_RWR}.

\section{Simulation}
In this section, we evaluate the operating characteristics of the proposed RWR method and compare to alternative methods by conducting simulation studies under two settings designed to reflect clinically relevant composite endpoints.
For each setting, we consider a two arm clinical trial with a sample size of 1200 participants, randomized in the 1:1 ratio to the treatment and control groups, denoted by $Z=1$ for treatment and $Z=0$ for control.
We examine the empirical type I error rate, statistical power, and confidence interval coverage of RWR. 
A two-sided significance level of 0.05 is used for hypothesis testing, and 95\% confidence intervals are constructed throughout. The empirical results are evaluated using 5,000 simulation replicates. All computations are performed in R \citep{Rprogramming}. 

\subsection{Multiple Time-to-event Endpoints}
\label{sec:TTE}
In the first setting, we consider four time-to-event endpoints with one fatal event and three non-fatal events, where the fatal event is prioritized over all non-fatal events and the non-fatal events are considered to have comparable clinical importance. To generate correlated event times, we adopt the Gumbel-Hougaard copula with exponential margins \citep{nelsen2006introduction}. Specifically, let 
$h_{D}(Z) = \lambda_D \exp(-\alpha_D Z)$ be the hazard rate for the fatal event and  
$h_{H_k}(Z) = \lambda_{H_k} (\alpha_{H_k} Z), k=\{1,2,3\}$ be the hazard rates for the three non-fatal events. Then the vector of time to four events (in days) $(D^{*},H^{*}_1,H^{*}_2,H^{*}_3)$ has the joint survival function:
$$
\mathbb{P}(D^*>y_1, H_{1}^*>y_2, H_{2}^*>y_3, H_{3}^*>y_4|Z) = \exp\left\{-\left[(h_{D}(Z) y_1)^\beta + \sum_{k=1}^3(h_{H_k}(Z) y_{k+1})^\beta \right]^{(1/\beta)}\right\},
$$ 
where $\beta \ge 1$ controls the correlation between any two endpoints (i.e., Kendall’s concordance equals $1-1/\beta$). We fix parameters $\lambda_D=0.0008,\allowbreak \bm\lambda_H=(\{\lambda_{H_k}\}_{k=1}^3)= \allowbreak (0.002,0.0015,0.001)$, $\beta=1.1, \alpha_D=0.2$. 
We consider four configurations of treatment effects on the non-fatal events that
$\bm\alpha_H = (\{\lambda_{H_k}\}_{k=1}^3) \in \{ (0.15,0.15,0.15), (0.2,0.15,0.1)
, (0.3,0.05,0.05), \allowbreak(0.05,0.05,0.3)\}$, representing different patterns of weak–strong effects across the non-fatal endpoints. It is possible to have a sample $i$ with $H_{ki}^*\geq D_i^*$, which indicates there is no $k$-th non-fatal event for $i$ due to censoring by death. Observed time to events is then obtained by performing administrative censoring after a fixed length of study, scheduled for 250 days to 1500 days, an accrual period of uniform entry of up to 200 days is assumed, and an independent dropout that follows $\text{Expn}(0.00016)$ distribution is also involved.
In addition to the proposed RWR, we consider the standard WR test applied under each of the six potential artificial orders of the three non-fatal events. Among these, we report WR-B (best case), defined as the standard WR test with the highest power across the six artificial orders, WR-W (worst case), defined as the standard WR test with the lowest power, and WR-R (random order), which applies the standard WR test using one randomly selected order from the six potential artificial ones. 
WR-R is included to represent an intermediate stance between  optimal and less favorable choices, and does not imply that the order would be chosen randomly in practice.
The log-rank test based on the traditional time-to-first-event analysis is also reported.

\begin{figure}
\centerline{\includegraphics[width=5in]{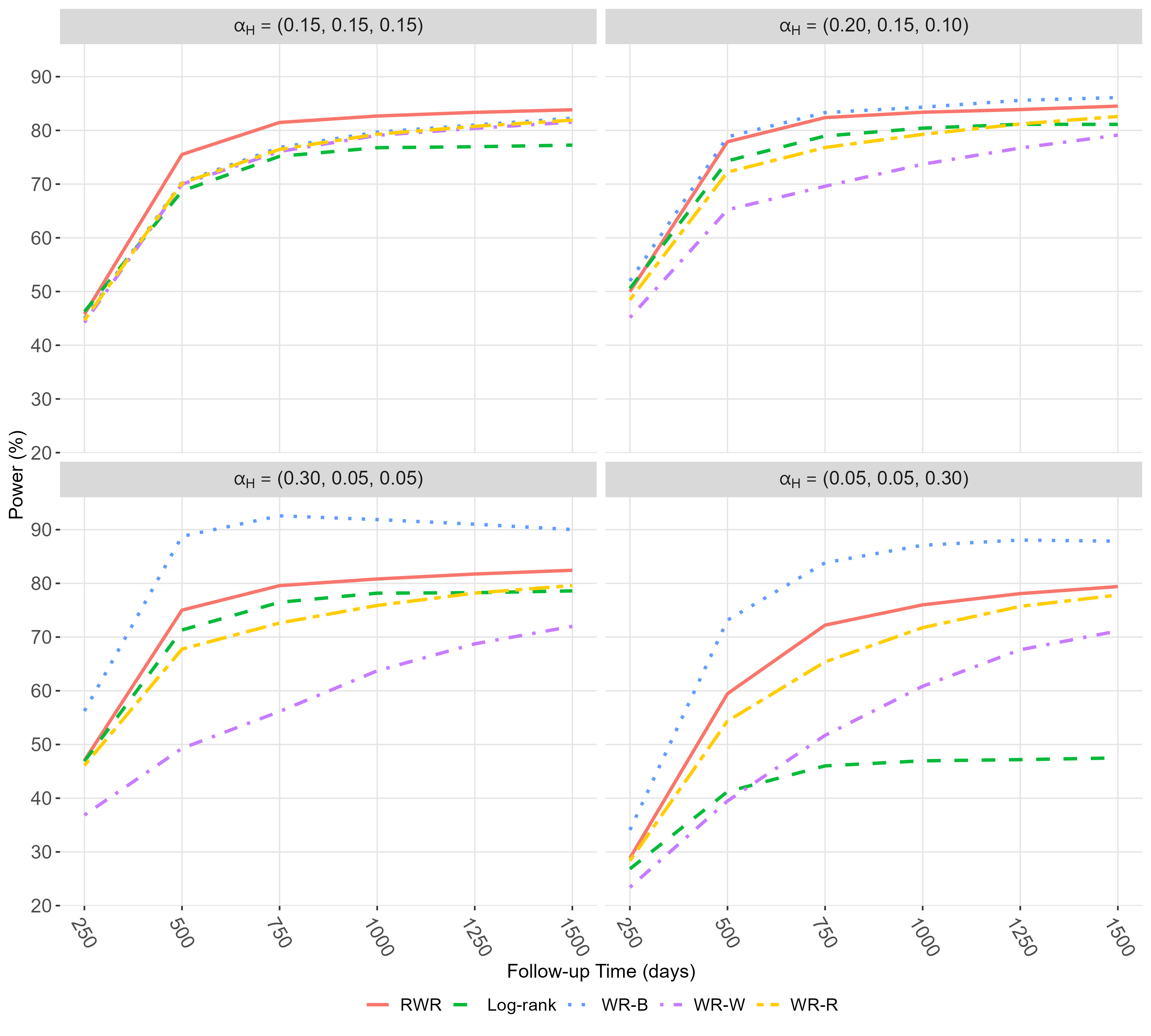}}
\caption{Empirical power of RWR, WR-B, WR-W, WR-R, and log-rank tests in simulation scenarios across different study durations for follow up and treatment effect magnitudes. \label{fig:TTE power}}
\end{figure}

The empirical power results are presented in Figure~\ref{fig:TTE power}. Across the four treatment effect configurations, several patterns emerge. First, under the balanced configuration ($\bm{\alpha}_{H}=(0.15,0.15,0.15)$), the differences in power across the included methods are relatively small. In this setting, WR-B, WR-W, WR-R, and log-rank tests yield broadly comparable performance, and RWR achieves power that is slightly higher than WR-B. This gain is likely due to the treatment effects on the three non-fatal endpoints being aligned, and placing them on equal priority allows RWR to exploit this consistent signal more effectively. 
Second, under the remaining three unbalanced configurations, the separation in empirical power among methods becomes pronounced. As expected, RWR delivers power that lies between that of WR-B and WR-W, reflecting its design of treating the three non-fatal endpoints symmetrically. Importantly, RWR consistently outperforms WR-R across all unbalanced configurations, demonstrating that aggregating information across all six rotations not only avoids dependence on an artificial ordering but also improves statistical efficiency compared with selecting a single rotation at random. This reinforces that the hybrid prioritization structure underlying RWR enhances both clinical interpretability and statistical power.
Third, comparing the two highly unbalanced configurations $\bm{\alpha}_{H}=(0.3,0.05,0.05)$ and $\bm{\alpha}_{H}=(0.05,0.05,0.3)$, all methods exhibit reduced power in the latter. This decline stems from the simulation design, where the third non-fatal endpoint has a smaller baseline hazard rate ($\lambda_{H_3}=0.001$) than the first two ($\lambda_{H_1}=0.002$, $\lambda_{H_2}=0.0015$), making it more difficult to detect a treatment effect of the same magnitude due to the reduced number of observable events prior to censoring. Although this general decrease affects all methods, the log-rank test is affected to the greatest extent. Since the log-rank test is based on time to the first event, the stronger treatment effect on the third non-fatal endpoint contributes less to the test statistic than an equally strong effect on the first non-fatal endpoint, simply because the latter is much more likely to occur first. Consequently, when the strongest signal is associated with an endpoint that is rarely the earliest event, the power of the log-rank test deteriorates more sharply than that of the win-based methods.

The empirical type I error rates for RWR under the null hypothesis ($\lambda_D = 0, \bm{\lambda}_H = \bm{0}$) range from 4.46\% to 5.06\% across different follow up durations. At the nominal level of $0.05$, RWR exhibits adequate type I error control, with observed rates falling well within the range expected from Monte Carlo variation.
In addition, the empirical coverage of the 95\% confidence intervals for RWR is reported in Table~\ref{tab:TTE CI}. Across all simulation configurations and follow up durations, the observed coverage probabilities are largely within the expected Monte Carlo variation range, with only a very slight deviation observed in one configuration (95.64\%). Overall, these results indicate that the proposed interval estimation procedure performs reliably in finite samples.

\begin{table}[htbp]
  \centering
  \caption{Empirical coverage of 95\% confidence interval of RWR in simulation scenarios across different study durations for follow up and treatment effect magnitudes. The expected range of empirical coverage with Monte Carlo variation under 5,000 replicates is 94.40\% to 95.60\% for the 95\% nominal level.}
    \begin{tabular}{lrrrr}
    \toprule
    \multicolumn{1}{c}{\multirow{2}[2]{*}{Follow up (days)}} & \multicolumn{4}{c}{$\bm \alpha_H$} \\
          & \multicolumn{1}{l}{(0.15,0.15,0.15)} & \multicolumn{1}{l}{(0.2,0.15,0.1)} & \multicolumn{1}{l}{(0.3,0.05,0.05)} & \multicolumn{1}{l}{(0.05,0.05,0.3)} \\
    \midrule
    250   & 95.18 & 95.30 & 95.02 & 94.98 \\
    500   & 95.06 & 95.12 & 95.64 & 94.74 \\
    750   & 95.00 & 95.46 & 95.48 & 95.08 \\
    1000  & 95.02 & 95.16 & 95.14 & 95.00 \\
    1250  & 94.90 & 95.04 & 95.32 & 94.98 \\
    1500  & 94.88 & 95.26 & 95.44 & 95.12 \\
    \bottomrule
    \end{tabular}%
  \label{tab:TTE CI}%
\end{table}%

\subsection{Fatal and Recurrent Events}
In the second setting, we consider scenarios involving one fatal event and one recurrent non-fatal event, with the fatal event prioritized above the recurrent event. For each participant, let $T_{j}$ denote the time to the $j$-th occurrence of the non-fatal event. We express the recurrent event trajectory in terms of gap times, defined by  
$
U_{1} = T_{1},
U_{2} = T_{2}- T_{1}, \ldots,
$
so that $U_{j}$ represents the waiting time between the $(j-1)$-th and $j$-th non-fatal events. To generate correlated event times, we consider a Gamma frailty model for $(D,U_1,U_2,\ldots,U_J)$, where $J$ is the maximum number of recurrent events per participant. Let $D|\xi,Z \sim \text{Expn}\left\{\lambda_D\exp(-\alpha_D Z)\xi \right\}$ and $U_{j}|\xi,Z \sim \text{Expn}\left\{\lambda_U\exp(-\alpha_{U_j} Z)\xi\right\}$, with $\xi \sim \text{Gamma}(\gamma^{-1},\gamma^{-1}),\gamma>0$. The $\gamma$ parameter controls the correlation among fatal and non-fatal recurrent events. We fix $\lambda_D=0.0008, \lambda_U=0.01$, $\gamma=0.2$.
We vary $\alpha_{D}$ from $0.025$ to $0.25$ in increments of $0.025$, and consider $J = 2, 3, 4$ under two treatment effect configurations for the gap times. In the homogeneous configuration, we set 
$
\alpha_{U_j} = \alpha_{D}, j = 1,\ldots, J,
$
so that the treatment effect on each gap time is identical to that on the fatal event. In the heterogeneous configuration, we let 
$
\alpha_{U_1} = \alpha_{D}, \alpha_{U_j} = 0, j = 2,\ldots, J,
$
so that only the first recurrence is directly affected by treatment, while subsequent gap times follow the same distribution across two treatment arms.
The observed event times are then obtained by imposing administrative censoring based on a study duration fixed to 1000 days with a uniform accrual over a 200-day enrollment window, and applying independent dropout that follows $\text{Expn}(0.00016)$ distribution.

While there are multiple ways to incorporate recurrent event information into a WS analysis, we follow the approach of \citet{mao2022recurrentevent} for illustration. Specifically, we include three summaries of the recurrent process, namely the number of recurrent events (NRE), the time to the first recurrence (FRT), and the time to the last recurrence (LRT). In the RWR analysis with the hybrid-prioritization structure, the FRT and LRT are placed on equal priority, with the NRE prioritized above them. 
Under the traditional fully prioritized structure, in contrast, it is typically difficult to include both the FRT and LRT simultaneously. Therefore, we consider WR-F (NRE prioritized over FRT) and WR-L (NRE prioritized over LRT) for comparison.

\begin{figure}
\centerline{\includegraphics[width=6in]{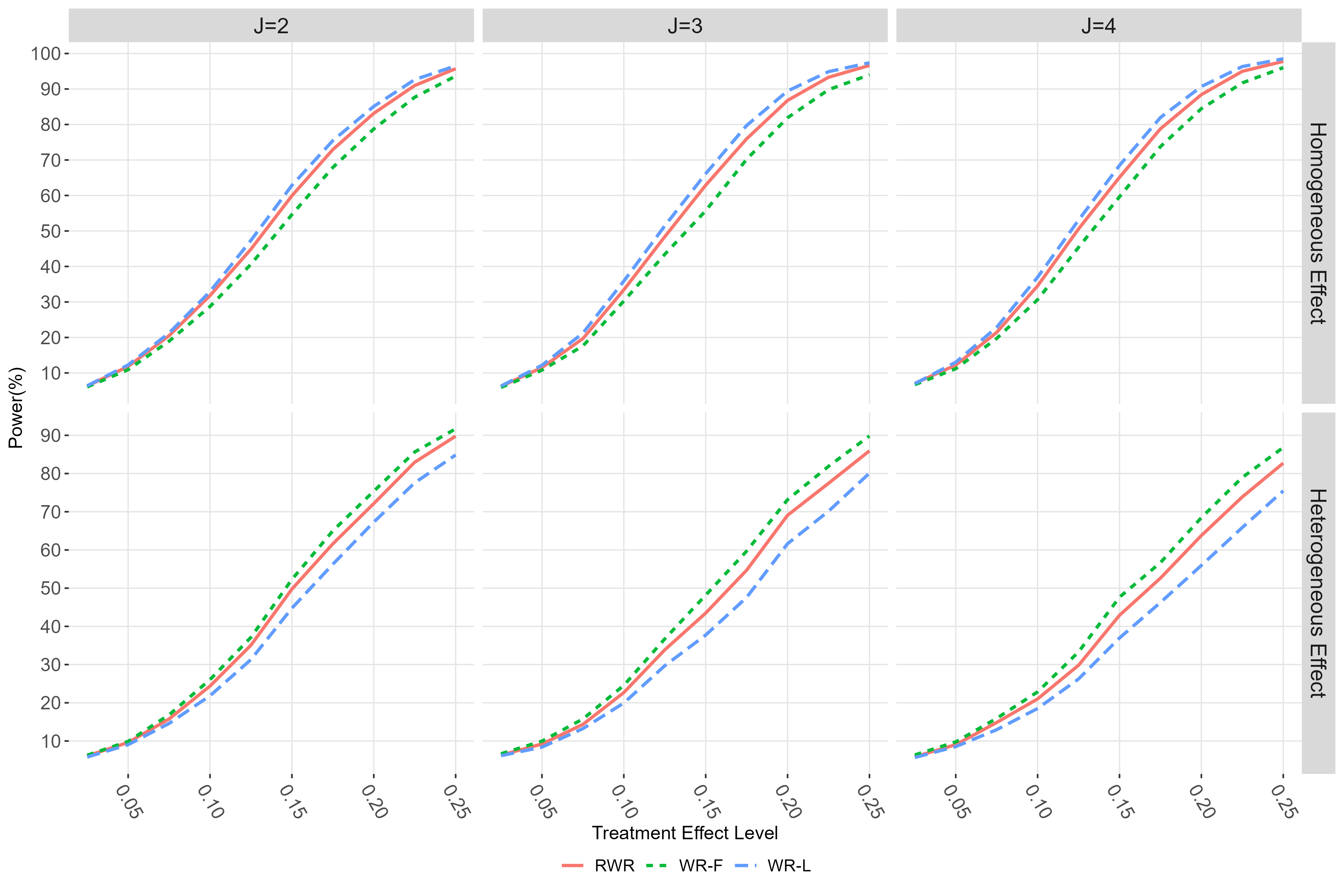}}
\caption{Empirical power of RWR, WR-F, and WR-L under homogeneous and heterogeneous treatment effects on recurrent event gap times, across treatment effect magnitudes and maximum numbers of recurrences ($J$). \label{fig:recurrent power}}
\end{figure}

The empirical power is presented in Figure~\ref{fig:recurrent power}. Across the simulation scenarios, several trends emerge. 
First, under the homogeneous treatment effect configuration for the gap times, WR-L achieves higher power than WR-F. This occurs because, when the treatment contantly affects every recurrence gap, the time to the last recurrence (if multiple events occur) reflects the cumulative effect across all gaps. As a result, the induced separation between treatment and control groups is larger for the LRT than for the FRT, making the WR-L yield higher empirical power. In contrast, under the heterogeneous treatment effect configuration, where only the first gap is influenced by treatment, the last recurrence time may aggregate several subsequent gaps that carry no treatment effect. This dilution makes the last recurrence comparison less informative than the first recurrence comparison, causing WR-F to outperform WR-L in this setting. The divergence between these two methods becomes more pronounced as $J$ increases, since larger values of $J$ allow the last event time to accumulate more gaps without treatment effect, further weakening the detectable signal for WR-L.
Second, across all simulation scenarios, the empirical power achieved by RWR consistently falls between that of WR-F and WR-L, which aligns with the construction of RWR, where FRT and LRT are included at equal priority. Notably, RWR tends to lie closer to the more powerful of WR-F and WR-L. This behavior mirrors the pattern observed in the multiple time to event setting in Section 3.1, whereby aggregating information across rotations enables RWR to outperform analyses based on random or artificial orderings and to more effectively capture the underlying treatment effect. 
In addition, the empirical coverage of the 95\% confidence intervals for RWR is reported in Table~\ref{tab:recurrent CI}. Most coverage estimates fall within the expected Monte Carlo variation range, with only a few very slight deviations (from 94.38\% to 95.76\%). Overall, the confidence interval procedure demonstrates satisfactory finite-sample performance in the recurrent-event setting.
Under the null setting ($\alpha_{D}=0$ and $\alpha_{U_j}=0$ for $j=1,2,\ldots,J$), the empirical type I error rates are 5.34\%, 4.86\%, and 5.26\% for $J=2,3,4$, respectively. All three values fall within the Monte Carlo variation range for 5,000 replicates (4.41\% to 5.64\%), indicating that the type I error is also well controlled.

\begin{table}[htbp]
  \centering
  \caption{Empirical coverage of 95\% confidence interval of RWR under homogeneous and heterogeneous treatment effects on recurrent event gap times, across treatment effect magnitudes and maximum numbers of recurrences ($J$). The expected range of empirical coverage with Monte Carlo variation under 5,000 replicates is 94.40\% to 95.60\% for the 95\% nominal level.}
    \begin{tabular}{lrrrrrr}
    \toprule
    \multicolumn{1}{c}{\multirow{2}[2]{*}{$\alpha_D$}} & \multicolumn{3}{c}{Constant Effect} & \multicolumn{3}{c}{Time-varying Effect} \\
          & \multicolumn{1}{c}{J=2} & \multicolumn{1}{c}{J=3} & \multicolumn{1}{c}{J=4} & \multicolumn{1}{c}{J=2} & \multicolumn{1}{c}{J=3} & \multicolumn{1}{c}{J=4} \\
    \midrule
    0.025 & 95.20 & 95.26 & 94.58 & 95.04 & 94.42 & 95.02 \\
    0.05  & 95.30 & 94.90 & 94.62 & 95.04 & 94.98 & 95.14 \\
    0.075 & 94.60 & 95.38 & 94.82 & 95.74 & 95.16 & 94.42 \\
    0.1   & 95.14 & 94.78 & 95.02 & 94.84 & 95.12 & 95.28 \\
    0.125 & 95.20 & 95.32 & 94.62 & 94.66 & 95.16 & 94.70 \\
    0.15  & 94.98 & 95.14 & 94.46 & 95.76 & 95.44 & 95.30 \\
    0.175 & 94.94 & 95.12 & 95.36 & 94.38 & 95.32 & 94.58 \\
    0.2   & 95.08 & 95.00 & 95.26 & 94.40 & 94.96 & 95.58 \\
    0.225 & 94.70 & 94.66 & 95.16 & 95.00 & 95.48 & 94.90 \\
    0.25  & 94.96 & 94.74 & 95.24 & 95.00 & 95.10 & 94.76 \\
    \bottomrule
    \end{tabular}%
  \label{tab:recurrent CI}%
\end{table}%

In sum, this simulation setting illustrates how RWR can be used to incorporate multiple summaries of recurrent event information in a more comprehensive manner. In this example, consider the ordering NRE, FRT and LRT, following the fatal endpoint. If any difference in FRT is considered meaningful (i.e., having zero threshold for the FRT block), whenever two participants reach the FRT comparison, they must have tied on NRE, either because neither experienced a non-fatal event or because they experienced the same number of events. In both situations, LRT contributes little additional information, because either no recurrence times are observed or the continuous FRT is highly likely to resolve the tie before LRT is reached. Under the RWR structure, placing FRT and LRT at equal priority not only aligns better with their comparable clinical importance but also enables both summaries to contribute effectively to the overall comparison.

\FloatBarrier
\section{Case Study}
In this section, we illustrate the proposed method using data from the Systolic Blood Pressure Intervention Trial (SPRINT) \citep{sprint2015randomized}. The SPRINT was a large, multicenter, randomized clinical trial that evaluated whether intensive systolic blood pressure control (targeting less than 120 mmHg) reduces cardiovascular morbidity and mortality compared to standard management (targeting less than 140 mmHg) among adults with elevated cardiovascular risk but without diabetes. A total of 9,361 participants were randomized across 102 clinical sites and followed for a median of 3.26 years. The primary composite endpoint is time to the first occurrence of myocardial infarction (MI), other acute coronary syndromes (Non-MI ACS), stroke, heart failure (HF), or cardiovascular death events.
In our case study, we reanalyze the primary composite endpoint from SPRINT using the proposed hybrid-prioritization structure. Consistent with clinical severity, time to cardiovascular death is prioritized above all non-fatal events. Among the four non-fatal components, we place MI, non-MI ACS, and stroke at equal priority, and prioritize this group above HF. 
This ordering prioritizes acute vascular events over HF events to illustrate an example, but could be ordered in the opposite orientation if HF events were the main priority of interest.
Following the study design, we perform stratification by clinic. One clinic with only a single participant is excluded due to its limited within stratum sample size, resulting in a study population of 9,360 participants across 101 clinics, and $w^{(s)}=1$ is employed for all 101 strata.

Our analysis with the proposed method results in an estimated RWR of $1.32$ (95\% confidence interval $[1.12, 1.56]$), and the corresponding two-sided hypothesis test yields a p-value of $0.0012$, indicating a statistically significant beneficial treatment effect at the $0.05$ significance level. To further assess the stability of the interval estimate, we also compute a bootstrap confidence interval using 10,000 bootstrap resamples, which yield the same interval $[1.12, 1.56]$ as that obtained from the derived asymptotic distribution. 
Using the same hybrid-prioritization structure, the RNB (1.42\%, [0.58\%, 2.27\%]) and RWO (1.03, [1.01, 1.05]) analyses provide similar evidence of a beneficial treatment effect, with bootstrap confidence intervals almost identical to those from asymptotic distributions.
These findings are consistent with the original SPRINT analysis based on the primary time-to-first-event composite endpoint using the Cox proportional hazards model, which reported a hazard ratio of $0.75$ for the intensive treatment group (95\% confidence interval $[0.64, 0.89]$; p-value $< 0.001$).
Although the direction and magnitude of the estimated treatment effects, as well as the hypothesis testing results, are similar between the RWR analysis and the original time-to-first-event analysis, the RWR measure offers additional insights and a distinct interpretation. By incorporating a hybrid-hierarchical structure that reflects the differential clinical importance of the events within the primary composite endpoint, RWR potentially better aligns the analysis with clinical relevance and provides a more nuanced summary of the treatment effect.

Using WR as an example, the decomposition matrix summarizing the proportions of wins, losses, and ties is presented in Table~\ref{tab:SPRINT decom}, which is recommended as part of a comprehensive WS analysis \citep{pocock2023win, buyse2025handbook}. Under the proposed hybrid-prioritization structure, the second block contains the MI, non-MI ACS, and stroke endpoints. Because these endpoints are treated on equal priority, the reported proportions for this block are obtained by aggregating the corresponding win, loss, and tie counts across all underlying rotations.
Specifically, let $n_{t,i}^{(k)}$, $n_{c,i}^{(k)}$, and $n_{0,i}^{(k)}$ denote the numbers of wins, losses, and ties, respectively, at the $i$-th endpoint in the $k$-th rotation for the second block, which contains three equally prioritized endpoints. In the SPRINT application, $i=2,3,4$ index the MI, non-MI ACS, and stroke endpoints under different orderings across the six rotations ($k=1,2,\ldots,6$).
The proportion of wins is calculated by aggregating the win counts across all six rotations and across the three endpoints in this block:
$$
\frac{\sum_{k=1}^6 \sum_{i=2}^4 n_{t,i}\ksup}
    {\sum_{k=1}^6 \left( n_{t,1}\ksup + n_{c,1}\ksup + n_{0,1}\ksup \right)},
$$
and the proportion of losses is computed analogously by replacing $n_{t,i}\ksup$ with $n_{c,i}\ksup$ in the numerator.  
Since the total number of ties after comparing these three endpoints remains unchanged across all rotations, the proportion of ties can be obtained directly as
$
\left. n_{0,4}^{(1)} \;\middle/ \left( n_{t,1}^{(1)} + n_{c,1}^{(1)} + n_{0,1}^{(1)} \right) \right.,
$
where the quantities from the first rotation are representative of all rotations.
In the stratified version, the aggregated win counts are defined as  
$n_{t,i}\ksup = \sum_{s=1}^{m} w^{(s)} n_{t,i}^{(k)(s)}$,  
where $n_{t,i}^{(k)(s)}$ is the number of wins at the $i$-th endpoint in the $k$-th rotation within the $s$-th stratum. In the SPRINT application, we have $m=101$ strata (clinics) and $w^{(s)}=1$ for all $s=1,2,\ldots,101$.
To further aid interpretation, we also report the block-level WR for each block, which summarizes the relative contribution of wins and losses at each hierarchical level to the overall RWR. As shown in Table~\ref{tab:SPRINT decom}, the second block, which comprises MI, non-MI ACS, and stroke endpoints, resolves the largest proportion of ties among the three blocks. The protective treatment effect is more pronounced in the first and third blocks, leading to higher block-level WR values for cardiovascular death (1.73) and HF (1.75) than for the second block (1.25). Overall, even after comparisons across all three blocks involving five endpoints, a substantial proportion of pairwise comparisons remain tied. This is likely attributable to the relatively low event rates in the SPRINT, which limit the number of informative comparisons available for distinguishing between treatment groups. However, it is also notable that, despite the relatively low event rates, the large sample size in the SPRINT ensures the absolute numbers of observed events remain adequate. As a result, the analysis still achieves sufficient power to detect the protective treatment effect, as reflected in both the RWR estimate and the corresponding hypothesis testing results. Additional details on the underlying six rotations used in the RWR analysis are provided in the Supporting Information~S1.

\begin{table}[htbp]
  \centering
  \caption{Block-level decomposition of wins, losses, and ties in the RWR analysis of the SPRINT. For the block containing MI, non-MI ACS, and stroke, the reported values reflect the equally-prioritized treatment of these three endpoints and are obtained by averaging the corresponding comparison results across all underlying rotations.}
    \begin{tabular}{lrrrr}
    \toprule
    Block & \multicolumn{1}{l}{Wins (\%)} & \multicolumn{1}{l}{Ties (\%)} & \multicolumn{1}{l}{Losses (\%)} & \multicolumn{1}{l}{Block-level WR} \\
    \midrule
    Cardiovascular Death & 1.21  & 98.09 & 0.70   & 1.73 \\
    MI, Non-MI ACS, Stroke & 3.84  & 91.19 & 3.07  & 1.25 \\
    HF    & 1.05  & 89.54 & 0.60   & 1.75 \\
    \bottomrule
    \end{tabular}%
  \label{tab:SPRINT decom}%
\end{table}%

\section{Discussion}
In this study, we develop a hybrid-prioritization method that extends the WS framework to simultaneously accommodate strictly prioritized and equally-prioritized endpoints, and provide statistical inference based on U-statistics, including extensions to the stratified analysis.
A few practical considerations may help guide the use of the hybrid-prioritization structure in applied analyses. 
First, the interpretation of the Rotation WS measures remains fully parallel to that of their underlying WS counterparts; the key distinction lies in the endpoint ordering, where the Rotation WS approach allows blocks of equally-prioritized endpoints to be incorporated explicitly. When all endpoints are strictly prioritized, the Rotation WS reduces immediately to the corresponding standard WS. Consistent with existing recommendations for WS analyses, we advocate reporting a decomposition matrix, similar to Table~\ref{tab:SPRINT decom}, that summarizes the proportions of wins, losses, and ties at each block, as this provides important insight into how treatment effects accumulate across the prioritized structure. For analyses involving equally-prioritized blocks, the primary decomposition matrix should reflect the hybrid structure by correctly aggregating endpoints within the same equally-prioritized block, while supplementary decomposition tables, such as those displaying results for each underlying rotation, may be included to offer additional detail.
Second, in the hybrid-prioritization structure, all permutations of endpoints within an equally-prioritized block are incorporated through the rotation set. Because the number of rotations grows factorially with the size of each equally-prioritized block, computational burden can increase rapidly as more endpoints are included in the equally-prioritized block. As a practical rule of thumb, we recommend limiting any single block to at most four or five endpoints, as a block of five endpoints already generates $5! = 120$ rotations, which can be computationally intensive. This recommendation is consistent with general principles for constructing composite endpoints in clinical trials, where the inclusion of each endpoint requires careful justification rather than convenience. It is also worth noting that multiple smaller equally-prioritized blocks are often more manageable. For example, while one block containing six endpoints would yield $6! = 720$ rotations, which is typically impractical, two blocks each containing three endpoints would produce only $3! \times 3! = 36$ rotations.
Third, the proposed hybrid-prioritization structure is intended to improve alignment between the pairwise comparison framework and the underlying clinical relevance. Although we include comparisons of empirical power with alternative methods in the simulation studies to illustrate the operating characteristics, the hybrid-prioritization strategy should not be viewed as a mechanism solely for improving statistical power. Rather, its primary purpose is to provide a principled way to reflect situations in which certain endpoints share comparable importance, thereby enhancing the clinical interpretability of WS analyses.

There are also potential extensions and directions that may be of interest for further investigation. 
First, similar to the standard WS framework, the censoring of time-to-event endpoints may influence the estimation of the treatment effect measures, and this carries over to the Rotation WS as well. Existing methods developed to mitigate the impact of censoring in WS analyses may therefore be adapted to the hybrid-prioritization structure. For example, the inverse probability of censoring–weighted approach \citep{dong2020inverse} can be incorporated into the Rotation WS, with the weights applied consistently across all underlying rotations.
Second, although the interpretation of the Rotation WS measures is parallel to that of the standard WS, the applicability of existing sample size formulas warrants further investigation. For example, the sample size formula for the WR developed by \cite{yu2022sample} relies on inputs such as the overall WR and the probability of ties to determine the required sample size for specified type~I and type~II error rates. While the Rotation WS measures share conceptual similarities with their standard counterparts, they aggregate information across multiple underlying rotations, leading to different covariance structures. Thus, although existing formulas may offer a starting point, directly substituting WR with RWR in such formulas may not be theoretically justified without additional work. Because simulation based sample size determination for pairwise comparison methods can be computationally intensive, developing analytic sample size procedures tailored to the Rotation WS framework represents another potential direction for further investigation.

\bibliographystyle{biom}
\bibliography{ref}%

\newpage

{\LARGE \textbf{Supporting Information} }
\appendix

\renewcommand{\thesection}{S\arabic{section}}
\renewcommand{\thesubsection}{S\arabic{section}.\arabic{subsection}}
% table number S1, S2, ...
\renewcommand{\thetable}{S\arabic{table}}
\setcounter{table}{0}

\section{Details of Underlying Rotations in the RWR Analysis of the SPRINT Case Study}
\label{apx:SPRINT}
The endpoint-level and overall WR values of six underlying rotations in the SPRINT case study is presented in Table~\ref{tab:apx SPRINT}. Across the six underlying rotations, the overall WR values are nearly identical and closely match the RWR estimate, indicating that the protective treatment effect is highly consistent regardless of the ordering imposed on the three endpoints of equal importance. Examining the endpoint-level WR values for MI, non-MI ACS, and stroke further explains this consistency, as their values are similar across all rotations, suggesting that the specific ordering of these endpoints has only a small impact on their contribution to the overall analysis. This stability is likely a result of the event structure in the SPRINT data. Among participants, 4.25\% experienced at least one of these three events, but only 0.29\% experienced two such events, and none experienced all three. Consequently, these endpoints tend to resolve ties for different pairs of participants rather than overlapping within the same comparisons, making the influence of their ordering across rotations relatively small.

\begin{table}[htbp]
  \centering
  \caption{Endpoint-level WR and overall WR values for the six underlying rotations in the RWR analysis of the SPRINT case study. The shaded rows correspond to the block in which MI, non-MI ACS, and stroke endpoints are treated on equal priority; in each rotation, these three endpoints are prioritized in different orders. Cells with gray, green, and blue background colors indicate the endpoint-level WR values for MI, non-MI ACS, and stroke endpoints, respectively.}
    \begin{tabular}{lrrrrrr}
    \toprule
    \multirow{2}[2]{*}{Endpoint} & \multicolumn{6}{c}{Rotation} \\
          & 1     & 2     & 3     & 4     & 5     & 6 \\
    \midrule
    Cardiovascular Death & 1.714 & 1.714 & 1.714 & 1.714 & 1.714 & 1.714 \\
    
    \multirow{3}[0]{*}{MI, Non-MI ACS, Stroke} 
          & \cellcolor{Color1} 1.392 & \cellcolor{Color1} 1.392 
          & \cellcolor{Color2} 1.327 & \cellcolor{Color2} 1.327 
          & \cellcolor{Color3} 0.795 & \cellcolor{Color3} 0.795 \\

          & \cellcolor{Color3} 0.791 & \cellcolor{Color2} 1.383 
          & \cellcolor{Color1} 1.426 & \cellcolor{Color3} 0.806 
          & \cellcolor{Color2} 1.339 & \cellcolor{Color1} 1.418 \\

          & \cellcolor{Color2} 1.399 & \cellcolor{Color3} 0.804 
          & \cellcolor{Color3} 0.804 & \cellcolor{Color1} 1.456 
          & \cellcolor{Color1} 1.456 & \cellcolor{Color2} 1.399 \\

    HF    & 1.753 & 1.753 & 1.753 & 1.753 & 1.753 & 1.753 \\
    Overall WR & 1.322 & 1.323 & 1.324 & 1.320 & 1.319 & 1.319 \\
    \bottomrule
    \end{tabular}%
  \label{tab:apx SPRINT}%
\end{table}%
\end{document}